\pdfoutput=1
\newif\ifFull
\Fullfalse
\ifFull
\documentclass[11pt]{article}
\usepackage{fullpage}
\else
\documentclass{acm_proc_article-sp}
\fi
\usepackage{graphicx}
\usepackage{cite}
\usepackage{url}
\usepackage{latexsym}
\usepackage{amssymb}
\usepackage{amsmath}
\usepackage{algorithm2e}
\usepackage{multirow}
\usepackage{array}
\usepackage{tabularx}

\ifFull

\fi

\begin{document}

\pagestyle{plain}
\def\thepage{\arabic{page}}
\thispagestyle{empty}

\title{Two-Phase Bicriterion Search for Finding \\ 
Fast and Efficient Electric Vehicle Routes}

\author{%
Michael T.~Goodrich\\
Dept.~of Computer Science \\
University of California, Irvine \\
\url{http://www.ics.uci.edu/~goodrich}
\and
Pawe{\l} Pszona \\
Dept.~of Computer Science \\
University of California, Irvine \\
\url{http://www.ics.uci.edu/~ppszona}
}

\date{}
\maketitle 

\begin{abstract}
The problem of finding an electric vehicle route that
optimizes both driving time and energy consumption can be modeled as 
a bicriterion path problem. 
Unfortunately, the problem of finding optimal bicriterion paths
is NP-complete.
This paper studies such problems
restricted to \emph{two-phase} paths, which correspond to
a common way people drive electric vehicles, where a driver
uses one driving style (say, minimizing driving time) 
at the beginning of a route
and another driving style (say, minimizing energy consumption) at the end.
We provide efficient polynomial-time
algorithms for finding optimal two-phase paths in bicriterion networks,
and we empirically verify the effectiveness 
of these algorithms for finding good electric vehicle driving routes
in the road networks of various U.S.~states.
In addition, we show how to incorporate charging stations
into these algorithms, in spite of the computational challenges
introduced by the negative energy consumption of such network vertices.

\noindent\textbf{Keywords:} road networks, electric vehicles,
shortest paths, bicriterion paths, 
NP-complete.
\end{abstract}

\section{Introduction}
\begin{figure}[htbp]
\begin{center}
\includegraphics[width=3.3in]{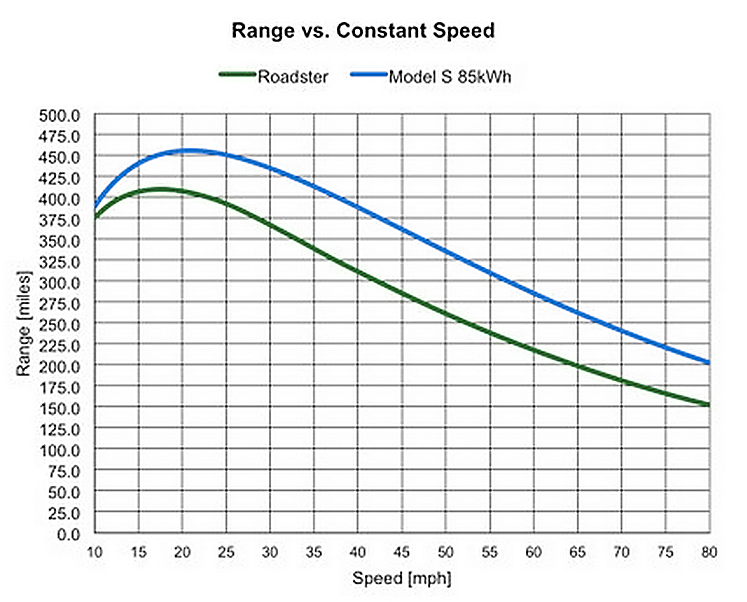}
\end{center}
\vspace*{-8pt}
\caption{Range versus speed for a Tesla Roadster and Tesla Model S
with 85 kWh battery~\cite{tesla}.}
\label{fig-tesla-range}

\begin{center}
\includegraphics[width=3.3in]{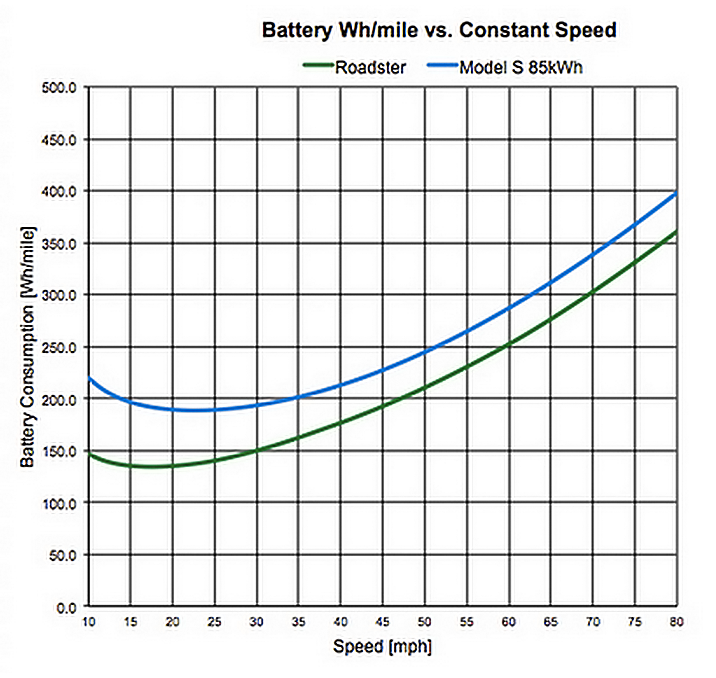}
\end{center}
\vspace*{-8pt}
\caption{Battery consumption per mile
for a Tesla Roadster and Tesla Model S 85kWh~\cite{tesla}.}
\label{fig-tesla-mph}
\end{figure}

Finding an optimal path for an electric vehicle (EV)
in a road network, from a given origin to a given destination, 
involves optimizing two criteria---driving time 
and energy consumption.
Unfortunately, these two criteria are usually in conflict, since people 
typically would like to minimize
driving time, but EVs are least efficient at high speeds.
(E.g., see Figures~\ref{fig-tesla-range} and~\ref{fig-tesla-mph}.)
Thus, planning good driving routes for EVs is
challenging~\cite{Franke201356,GrahamRowe2012140}, 
leading some to refer to the stress of dealing with the
restricted driving distances imposed by battery capacities
as ``range anxiety''~\cite{APPS474}.
To help electric vehicle owners deal with range anxiety, therefore,
it would be ideal if
GIS route-planning systems could quickly provide electric vehicle
owners with routes that optimize a set of preferred 
trade-offs for time and energy,
based on the energy-usage characteristics and 
the battery capacity of their vehicle.

\subsection{Modeling EV Route Planning}
This electric-vehicle route-planning problem can be modeled as a 
\emph{bicriterion path optimization} problem~\cite{h-mcdmt-80}
(which is also known as the \emph{resource constrained shortest path} 
problem~\cite{mz-rcsp-00}), where
one is given a directed graph, $G=(V,E)$,
such that each edge, $e \in G$, has a weight, $w(e)$, 
that is a pair of integers, $(x,y)$,
such that cost of traversing $e$ uses $x$ units of one type and $y$ units
of a second type. For instance, in a road network calibrated for a certain
electric vehicle, a given edge, $e$, might have a 
weight, $w(e)=(75,\,304)$, which
indicates that driving at a given speed (say, 60 mph) will require 75 seconds
and consume 304 Wh to traverse $e$.

The graph $G$ is allowed to contain
parallel edges, that is, multiple edges having
the same origin and destination, $v$ and $w$, 
so as to represent different ways of going from $v$ to $w$. For example, 
one edge, $e_1=(v,w)$, could represent a traversal
from $v$ to $w$
at 60 mph, another edge, $e_2=(v,w)$, could representing a traversal 
from $v$ to $w$ at 55 mph,
and yet another edge, $e_3=(v,w)$, could represent a traversal at 65 mph.

\begin{figure}[htb]
\vspace*{-12pt}
\begin{center}
\includegraphics[width=3.3in, trim = 0.3in 0.3in 3.5in 0.3in, clip]{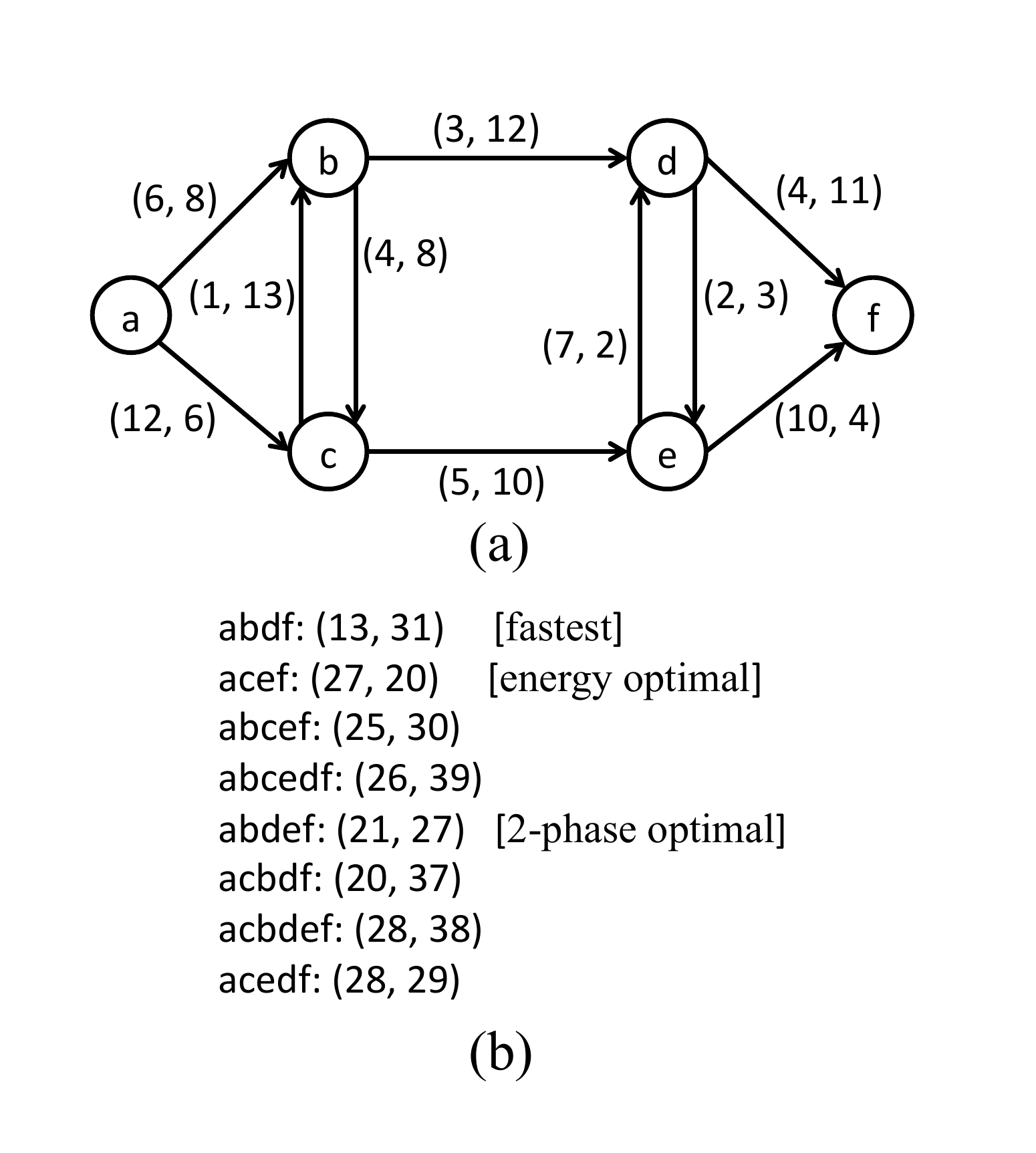}
\end{center}
\vspace*{-28pt}
\caption{An instance of the bicriterion path problem. (a) A network with 
(driving-time,~energy-consumption) edge weights;
(b) All the 
paths in the graph and their respective weights. We highlight 3 interesting
path weights.}
\label{fig-two-phase}
\end{figure}

For a path, $P=(e_1,e_2,\ldots,e_k)$, in $G$, 
whose edges have respective weights,
$(x_1,y_1)$, $\ldots$, $(x_k,y_k)$, the weight, $w(P)$, of $P$, is defined as
\[
w(P) = \left( \sum_{i=1}^k x_i\,,\,\,\, \sum_{i=1}^k y_i \right).
\]
Given a starting vertex, $s$, and a target vertex, $t$, and two 
integer parameters,
$X$ and $Y$, the 
\emph{bicriterion path problem} is to find a path, $P$, in $G$, from $s$
to $t$, such that $w(P)=(x,y)$ with $x\le X$ and $y\le Y$.
(See Figure~\ref{fig-two-phase}.)
Unfortunately, as we review below, the bicriterion path problem 
is NP-complete.

The bicriterion path problem has a rich history, and several heuristic
and approximation
algorithms have been proposed to solve it 
(e.g., see~\cite{h-mcdmt-80,h-asrsp-92,Henig1986281,Modesti1998495,%
mz-rcsp-00,Mote199181,NamoradoClimaco1982399,Skriver2000507}).
Rather than take a heuristic or approximate approach, however, we are interested
here in reformulating the problem so as to simultaneously achieve the following
goals:
\begin{itemize}
\item
The formulation should capture the way people drive electronic vehicles 
in the real world.
\item
This formulation should be solvable
in (strongly) polynomial time, ideally, with the same asymptotic 
worst-case running time
needed to solve a single-criterion shortest path problem.
\end{itemize}

\subsection{Our Results}
In this paper, we show that one can, indeed, achieve both of the
above goals by using a formulation we call
the \emph{two-phase bicriterion path problem}.
In a \emph{two-phase path}, $P$, we
traverse the first part of $P$ according to one driving
style and we traverse the remainder of $P$ according to a second driving style.
For instance, we might begin an electric vehicle route optimizing 
primarily for driving time but finish this route optimizing primarily
for energy consumption, which is a common way electric vehicles are driven
in the real world (e.g., see~\cite{Franke201356,GrahamRowe2012140}).
We provide a general mathematical framework for the two-phase bicriterion
path problem and
we show how to find such paths in a network of $n$ vertices and $m$
edges in $O(n\log n + m)$ time, 
if edge weights are pairs of non-negative integers,
and in $O(nm)$ time otherwise.
In addition, we show to extend our algorithms to incorporate charging
stations in the network, with similar running times.
We include an experimental validation of our algorithms using Tiger/Line
USA road network data, showing that our algorithms are effective both in terms of
their running times and in terms of the quality of the solutions that they
find.

\subsection{Additional Related Work}
In ACM SIGSPATIAL GIS '13,
Baum {\it et al.}~\cite{Baum:2013} describe an algorithm for finding 
energy-optimal routes for electric vehicles, based on a variant of Dijkstra's
shortest path algorithm. 
They contrast the paths
their algorithm finds with shortest travel time and shortest distance paths,
showing that the paths found by their algorithm are significantly more energy
efficient.
In addition to this work, the problem of finding
energy-optimal paths for electric vehicles
is also studied by 
Artmeier {\it et al.}~\cite{artmeier-10},
Eisner {\it et al.}~\cite{eisner2011optimal},
and
Sachenbacher {\it et al.}~\cite{sachenbacher2011efficient}.
Unfortunately, 
these energy-optimal paths are not that practically useful
for typical drivers of electric vehicles, who
care more about quickly reaching their
destinations (while not depleting their batteries) 
than they do about minimizing overall energy consumption
(e.g., see~\cite{Franke201356,GrahamRowe2012140}).
For instance, as shown in 
Figures~\ref{fig-tesla-range} and~\ref{fig-tesla-mph},
in a Tesla Roadster or Model S 85kWh, 
a driver achieves optimal
energy efficiency on level ground by maintaining a constant speed of 15 to 20
mph, which is unrealistic for real-world road trips.
Thus, we feel it is more productive to provide algorithms that can find routes
with small travel times that also conserve sufficient energy to avoid
fully depleting a vehicle's battery (if possible), 
which motivates studying electric vehicle
route planning as a bicriterion path problem.

We are not familiar with any prior work on finding optimal two-phase
bicriterion paths, but 
there are well-known algorithms 
for finding single-phase paths and
for enumerating all Pareto optimal bicriterion paths.
We review these classic results in the next section.

Bidirectional shortest-path algorithms have been used
as an approach to speedup 
shortest path searching~\cite{gkw-alenex,Righini2006255},
but, to our knowledge, these have not been applied in the way we are doing
bidirectional search for finding optimal two-phase shortest paths.
In addition, Storandt~\cite{Storandt:2012} studies EV route planning
taking into account charging stations, but not in the same way 
that we incorporate charging stations into two-phase routes.

\section{The Complexity of Bicriterion Path Finding}
\label{pseudo_polynomial_sec}
We begin by reviewing known results for
the bicriterion path problem, absent of the two-phase path formulation,
including
that finding bicriterion shortest paths is NP-complete, but there
is a pseudo-polynomial time algorithm for finding bicriterion paths, which
can be very slow in practice.

\subsection{Bicriterion Path Finding is NP-Complete}
The bicriterion path problem
is NP-complete, even if the values in the weight pairs are all positive
integers
(e.g., see~\cite{arkin1991bicriteria,gj-cigtn-79}).
For instance, there is a simple polynomial-time
reduction from the Partition problem,
where one is given a set, $A$, of $n$ positive numbers,
$A=\{a_1,a_2,\ldots,a_n\}$, and asked if there is a subset,
$B\subset A$, such that $\sum_{a_i\in B} a_i = \sum_{a_i\in A-B} a_i$.
To reduce this to the bicriterion path problem, 
let the set of vertices be $V=\{v_1,v_2,\ldots,v_{n+1}\}$,
and, for each $v_i$, $i=1,\ldots,n$, create two edges,
$e_{i,1}=(v_i,v_{i+1})$ and $e_{i,2}=(v_i,v_{i+1})$,
such that $w(e_{i,1})=(1+a_i,\,1)$ and $w(e_{i,2})=(1,\,1+a_i)$.
Let $h=(\sum_{i=1}^n a_i)/2$,
and define this instance of the bicriterion path problem to ask if
there is a path, $P$, from $v_1$ to $v_{n+1}$, with weight 
$w(P)=(x,y)$ such that $x\le n+h$ and $y\le n+h$.
This instance of the bicriterion path problem has a solution if and
only if there is a solution to the Partition problem.

\subsection{A Pseudo-Polynomial Time Algorithm}
As with the Partition problem, there is a pseudo-polynomial time
algorithm for the bicriterion path problem 
(e.g., see~\cite{h-mcdmt-80,h-asrsp-92}).
Recall that the input to this problem is an $n$-vertex graph, $G$, with integer
weight pairs stored at its $m$ edges (and assume for now that 
none of these values are negative), together with parameters $X$ and $Y$.
In this pseudo-polynomial time algorithm,
which we call the ``vertex-labeling'' algorithm,
we store at each vertex, $v$, a set of pairs,
$(x,y)$, such that there is a path, $P$, from $s$ to $v$ with weight $(x,y)$. 
We store such a pair, $(x,y)$, at $v$, if we have discovered
a path with this weight and only if, at this point
in the algorithm, there is no
other discovered weight pair, $(x',y')$, with $x'<x$ and $y'<y$, 
for a path from $s$ to $v$.

Initially, we store ${(0,0)}$ at $s$ and we store $\emptyset$ at
every other vertex in $G$.
Next, for a sequence of iterations, we perform a \emph{relaxation} for each
edge, $e=(v,w)$, in $G$, with $w(e)=(x,y)$, 
such that, for each pair, $(x',y')$, stored at $v$, we add 
$(x+x',\,y+y')$ to $w$, provided there is no pair, $(x'',y'')$,
already stored at $w$, such that $x''\le x+x'$ and $y''\le y+y'$.
Moreover, if we add such a pair $(x+x',\,y+y')$ to $w$, then we
remove each pair, $(x'',y'')$, from $w$ such that
$x''> x+x'$ and $y''> y+y'$.
The algorithm completes when an iteration causes
no label updates, at which point we then test if there is a pair,
$(x,y)$, stored at the target vertex, $t$, such that $x\le X$ and $y\le Y$.

If we let $N$ denote the maximum
value of a sum of $x$-values or $y$-values
along a path in $G$, then
the running time of this algorithm
is $O(nmN)$, because each iteration takes at most $O(mN)$ time and there
can be at most $O(n)$ iterations (since there can be no negative-weight cycles).
Because $N$ can be very large, this is only a pseudo-polynomial time
algorithm.
In practice, this algorithm can be quite inefficient; for instance,
in a road network, $G$, for an electric vehicle, $N$ could be 
the number of seconds in 
the maximum duration of a trip in $G$ or the capacity of the battery
measured in Wh.

\subsection{Battery Capacities and Charging Stations}
Although the above algorithm is not very efficient,
we can nevertheless modify it
to work for electric vehicle routes, taking into
consideration battery capacities and the existence of charging
stations.
Here, we assume that each edge weight $w(e)=(x,y)$, where $x$ is the
time to traverse the edge (at a speed associated with the edge $e$)
and $y$ is the energy consumed by this traversal.
We also assume that the vehicle starts its journey from the start
vertex, $s$, with a fully charged battery.

A charging station can be modeled as a vertex that has a
self-loop with a weight $(x,y)$ having a positive $x$ value and negative $y$.
There may be other edges in the graph with negative $y$-values, as
well, such as a stretch of road that goes sufficiently downhill to
allow net battery charging through regenerative braking.

We store at each vertex a collection of $(x,y)$ values
corresponding to the driving time, $x$, and net energy consumption,
$y$, along some path starting from the start vertex, $s$.
We modify the above vertex-labeling 
algorithm, however, to disallow storing an $(x,y)$ pair with
a negative $y$ value, 
since we assume our vehicle begins with a fully charged
battery, and it is not possible to store more energy 
in a battery after it is fully charged.

Similarly, we assume we know the capacity, $C$, for the vehicle's
battery. If we ever consider a weight pair, $(x,y)$, for an $s$-to-$w$ path,
such that $y>C$, then we discard this pair and do not add it to
the label set for $w$.
Such a pair $(x,y)$ corresponds to a path that would fully discharge 
the battery;
hence, attempting to traverse this path would cause the vehicle to
stop functioning and it would not reach its destination.

Making these modifications allows the vertex-labeling algorithm to be
adapted to an environment for planning the route of an electric
vehicle, including consideration of its battery capacity, the fact
that its battery cannot hold more than a full charge, and removal
of paths that would require too much energy to traverse.
These modifications do not improve its asymptotic running time,
however, which becomes $O(nmN^2)$, where $N$ is the largest route
duration or the battery capacity, since each iteration takes
$O(mN)$ time and there can be at most $O(nN)$ iterations (given
our restrictions based on the battery capacity).

\subsection{Drawbacks}
In addition to its inefficiency, the 
vertex-labeling algorithm
might find an
optimal path 
that could be difficult to actually drive in practice.
For instance, it could involve many alternations between various styles
of driving, such as ``drive the speed limit''
and ``drive 10 mph below the speed limit.'' 
In addition, it could involve several detours, for instance, asking a driver
to systematically 
get on and off a limited-access high-speed highway. Such detours
are distracting and difficult to follow, of course, but 
they could also be expensive, if that limited-access highway were a
toll road.
Thus, implementing 
the so-called ``optimal'' path that this algorithm produces might
require an onboard GPS system to constantly be barking out strange
orders to the driver, which, unless the driver enjoys
road rallies, could be difficult and annoying to follow.
Clearly, we prefer a formulation of the bicriterion path problem
that would better match the ways people drive in practice.

\section{Linear Utility Functions}
\label{sec:modes}
\begin{figure}[hbt]
\begin{center}
\includegraphics[width=3.5in, trim = 0.5in 3.1in 3.1in 1in, clip]{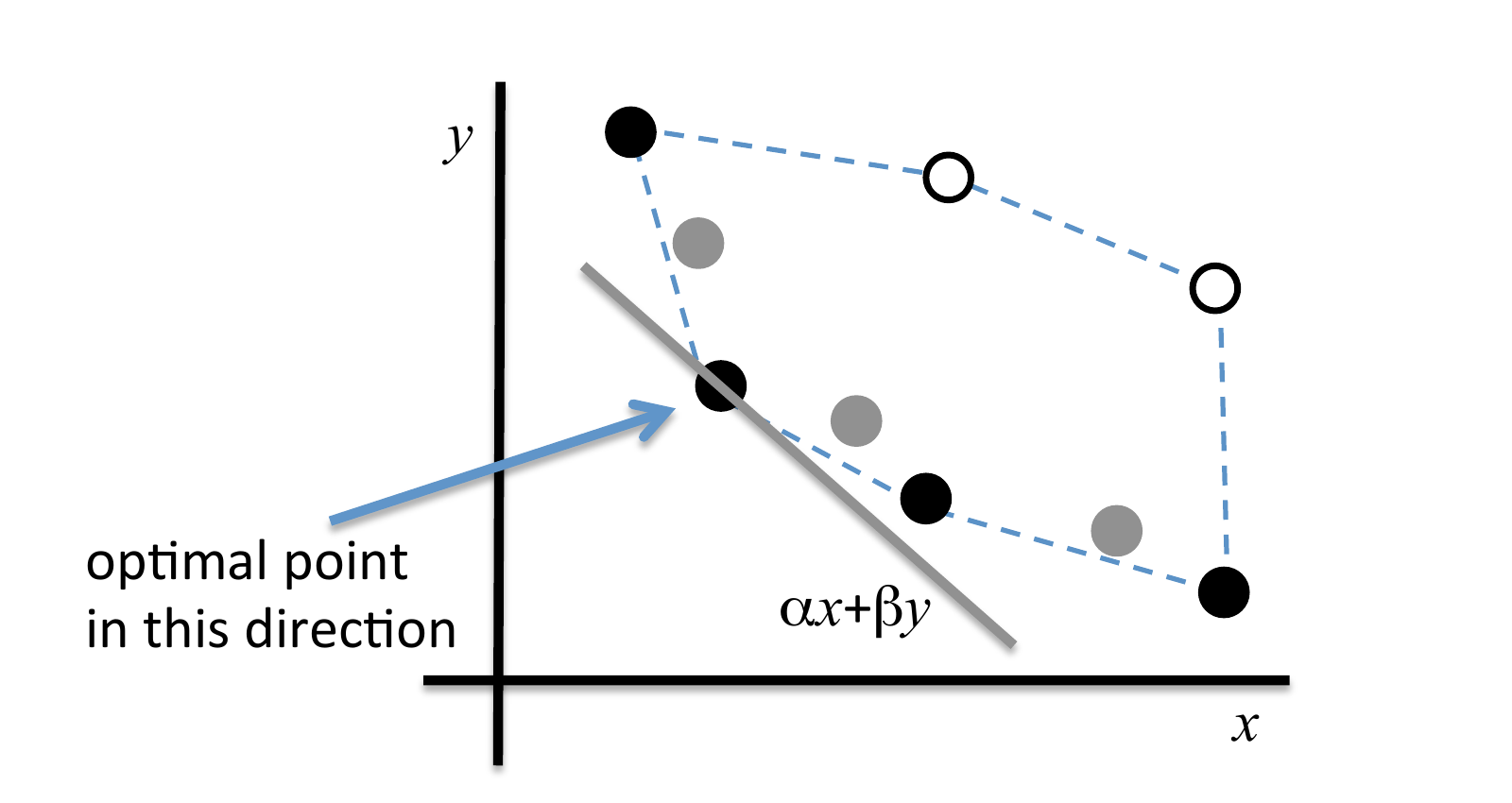}
\end{center}
\vspace*{-8pt}
\caption{Sample $(x,y)$ points that correspond to the weights of paths 
in a bicriterion network. The solid points could potentially be 
found by a linear optimization algorithm 
using an $(\alpha,\beta)$ preference pair, as they are on the convex
hull of the set of $(x,y)$ points, shown dashed.
The gray points are Pareto-optimal points (that is, not dominated
by any other point), but they would not be found
by an algorithm that searches for optimal paths based on linear utility
functions and preference pairs.
The empty points are not Pareto optimal; hence, they should not
be returned as options from a bicriterion optimization algorithm.}
\label{fig:hull}
\end{figure}

Fortunately,
there is a more natural and efficient algorithm for finding good bicriterion
paths, by using linear utility functions 
(e.g., see~\cite{Henig1986281,mz-rcsp-00,Modesti1998495}).
Suppose we are given a directed
network, $G$, together with pairs, $(x,y)$, defined
for each edge in $G$.
Formally, we define a linear utility function in terms
of a \emph{preference pair}, $(\alpha,\beta)$, of non-negative real numbers.
A path $P$, from $s$ to $t$, in $G$, is \emph{optimal} for 
a preference pair $(\alpha,\beta)$ 
if it minimizes the cost, $C_{\alpha,\beta}(P)$, of $P=(e_1,e_2,\ldots,e_k)$, 
with $w(e_i)=(x_i,y_i)$, 
\[
C_{\alpha,\beta}(P) = \sum_{i=1}^k (\alpha x_i + \beta y_i),
\]
taken over all possible paths from $s$ to $t$ in $G$ (that is, $k$ is
a free variable and we do not limit the number of edges in $P$).
For example, using the preference pair $(1,\, 0.01)$, for edge weights defined by
pairs of driving times in seconds and energy consumption in watt-hours,
would imply a driving style that tends
to emphasize driving time over energy consumption.
Note that we can also write this cost for a path, $P$, as two global sums,
\[
C_{\alpha,\beta}(P) = \sum_{i=1}^k \alpha x_i \,+\, \sum_{i=1}^k \beta y_i,
\]
which implies that we can visualize this optimization as 
that of finding a vertex
on the convex hull of $(x,y)$ points for the weights of $s$-$t$ paths in $G$,
in a direction determined by $\alpha$ and $\beta$.
Moreover, this algorithm cannot find $(x,y)$ points that are not on the 
convex hull.
(See Figure~\ref{fig:hull}.)

If the $\alpha x_i+\beta y_i$ values for the edges
in $G$ are all non-negative, then an optimal 
$s$-to-$t$ path, for any preference pair, $(\alpha,\beta)$, 
can be found using a standard single-source shortest
path algorithm~\cite{Henig1986281},
which runs in $O(n\log n+m)$ time,
where $n$ is the number of vertices in $G$ and $m$ is the number of edges,
by an implementation of Dijkstra's algorithm (e.g., see~\cite{clrs-ia-01}).
Otherwise, such a path can be found in $O(nm)$ time, by the 
Bellman-Ford algorithm (e.g., see~\cite{clrs-ia-01}).
Indeed, for any vertex, $v$, and a given preference pair, $(\alpha,\beta)$,
we can use these algorithms to find the tree 
defined by the union of all $(\alpha,\beta)$-optimal paths 
in $G$ that emanate out from $v$, or are directed into $v$,
in these same time bounds.
(Note that we may allow such paths to include
self-loops at charging stations a finite number of times, so that the 
topology of their union is still essentially a tree.)

\section{Two-Phase Bicriterion Paths}
\label{two_phase_sec}
Restriction to finding a route 
optimizing a single linear utility function, as described above,
may be too constraining. Because it misses $(x,y)$ pairs that
are not on the convex hull,
if we are planning a route from a source, $s$, to a target,
$t$, 
there might be fast and efficient $s$-to-$t$ path,
that is missed, since
a path minimizing driving time might run out of energy before
reaching $t$, while a route minimizing energy consumption might be
needlessly slow.
(See, for example, Figure~\ref{fig-two-phase}.)
Thus, it would be desirable to consider routes that include a transition
from one linear utility function to another at some point,
such as a route that optimizes driving time in the beginning
of the route and switches to optimizing energy consumption at the end,
so as to reach the target vertex quickly without fully discharging
the battery.

Suppose we are given two preference pairs, 
$(\alpha_1,\beta_1)$ and $(\alpha_2,\beta_2)$.
For example, we might have 
$(\alpha_1,\beta_1)=(1,0.1)$, which emphasizes
driving time, and $(\alpha_2,\beta_2)=(0.1,1)$, 
which emphasizes energy consumption.
A path, $P$, from $s$ to $t$ is a \emph{two-phase} path 
for $(\alpha_1,\beta_1)$ and $(\alpha_2,\beta_2)$
if there is a vertex, $v$, in $P$, such that 
we can divide $P$ into the path, $P_1$, from $s$ to $v$, and the
path, $P_2$, from $v$ to $t$, so that 
$P_1$ is an optimal $s$-to-$v$ path
for the preference pair $(\alpha_1,\beta_1)$ and
$P_2$ is an optimal $v$-to-$t$ path
for the preference pair $(\alpha_2,\beta_2)$.
(For example, in Figure~\ref{fig-two-phase},
the path $abdef$ is a composition of a
time-optimal path from $a$ to $d$ and an energy-optimal path from
$d$ to $f$, and this would be a two-phase optimal path for a battery capacity
from 27 to 30, inclusive.)
As a boundary case,
we allow the vertex $v$ to be equal to $s$ or $t$, 
so that a single-phase path is just a special case of 
a two-phase path.

\subsection{Finding Two-Phase Paths}
In this section, we describe our polynomial-time algorithm for
finding an optimal two-phase path from a source, $s$, to a target, $t$,
in a graph, $G$, with bicriterion weights on its edges.
We describe an algorithm that can search for two-phase paths based on
optimizing two out of $c$ given preference pairs.
Suppose, then, that we are given $c$ preference pairs,
$(\alpha_1,\beta_1), (\alpha_2,\beta_2), \ldots, (\alpha_c,\beta_c)$.

\begin{enumerate}
\item
For each
preference pair, $(\alpha_i,\beta_i)$, use the algorithm of
Section~\ref{sec:modes} to find the tree, $T^{\rm out}_{s,i}$, that is the
union, for all $v$ in $G$,
of the optimal $s$-to-$v$ paths in $G$ for the pair 
$(\alpha_i,\beta_i)$.
With each node, $v$, store the bicriterion weight, $(x,y)^{\rm out}_{i}$,
of the $s$-to-$v$ path in $T^{\rm out}_{s,i}$.

\item
For each
preference pair, $(\alpha_j,\beta_j)$, use the (reverse) algorithm of
Section~\ref{sec:modes} to find the tree, $T^{\rm in}_{t,j}$, that is the
union, for all $v$ in $G$,
of the optimal $v$-to-$t$ paths in $G$ for the pair
$(\alpha_j,\beta_j)$.
With each node, $v$, store the bicriterion weight, $(x,y)^{\rm in}_{j}$,
of the $s$-to-$v$ path in $T^{\rm in}_{t,j}$.
\item
For each node $v$ in $G$, and each pair of indices
$i,j=1,2,\ldots,c$, 
compute the score 
\[
(x,y)^v_{i,j} = (x,y)^{\rm out}_i + (x,y)^{\rm in}_j,
\]
for performing a transition from preference pair $(\alpha_i,\beta_i)$
to $(\alpha_j,\beta_j)$ at $v$,
where ``$+$'' is component-wise addition.
\item
Search all the $(x,y)^v_{i,j}$ values, including values $(x,y)^v_{i,i}$,
to find an optimal $(x,y)$
pair according
to the user's specified optimization
goals, such as $x\le X$ and $y\le Y$, for some $X$ and $Y$.
\end{enumerate}

\begin{figure}[hbt]
\begin{center}
\includegraphics[width=3.3in, trim = 0.7in 3.2in 0.7in 0.5in, clip]{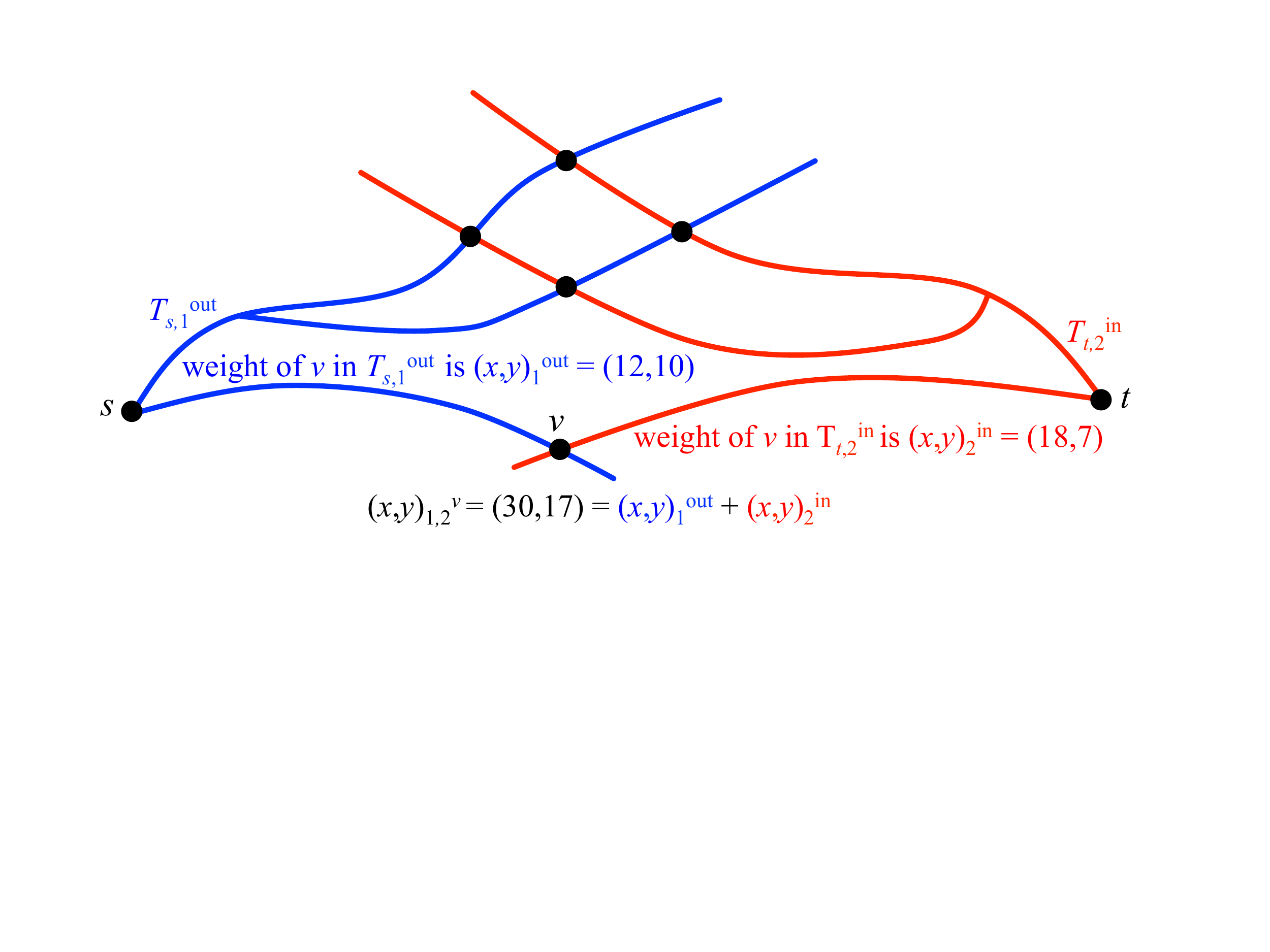}
\end{center}
\vspace*{-8pt}
\caption{Schematic illustration of the two-phase polynomial-time algorithm.}
\label{fig:mixing}
\end{figure}

We give a schematic illustration of this algorithm in Figure~\ref{fig:mixing}.

\begin{figure}[htb]
\begin{center}
\includegraphics[width=3.3in, trim = 2.0in 1.5in 2.5in 0.5in, clip]{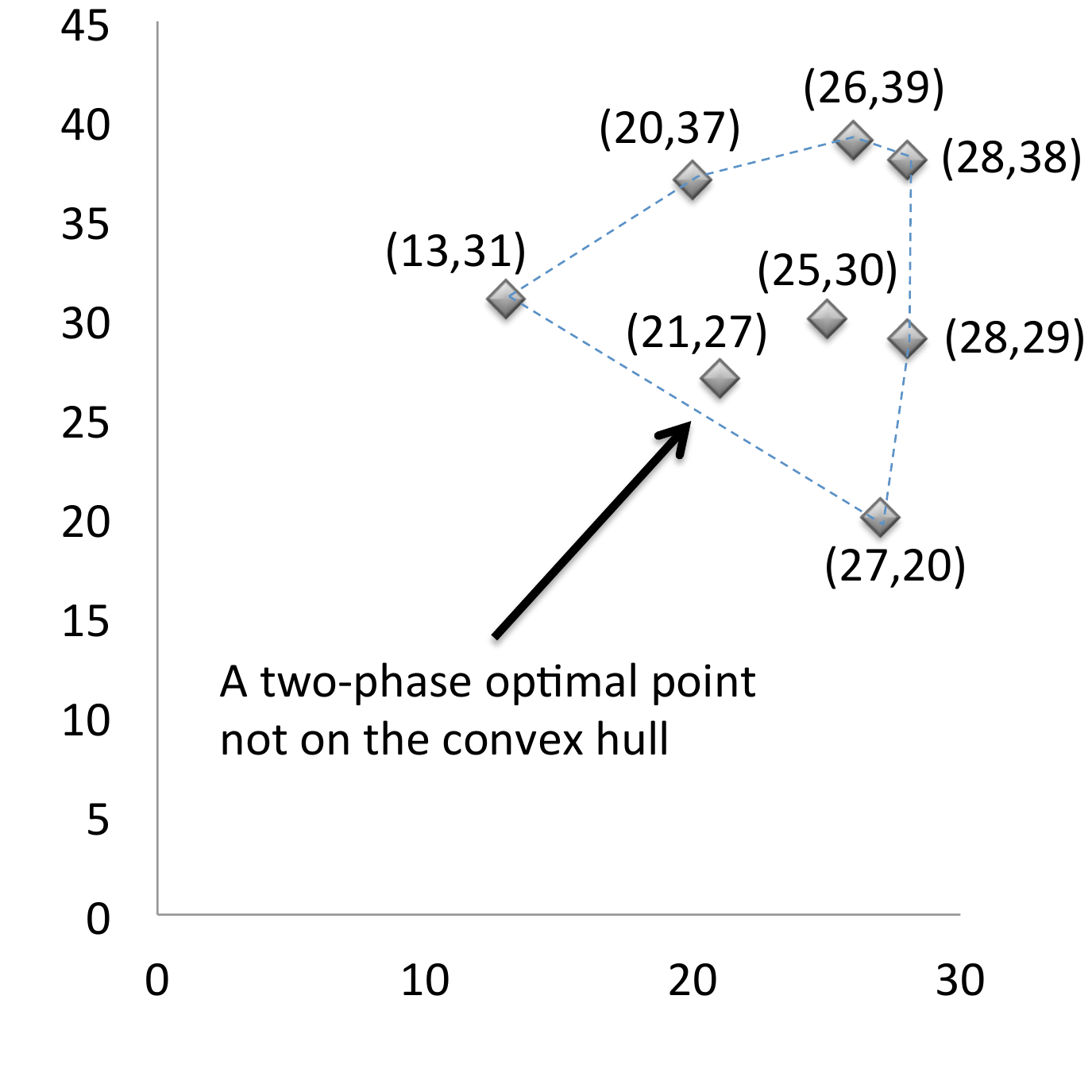}
\end{center}
\vspace*{-28pt}
\caption{A plot of the different weight pairs for $a$-to-$f$ paths in the network
of Figure~\protect{\ref{fig-two-phase}}. Note that the weight pair, $(21,27)$,
for the path $abdef$,
is not on the convex hull, shown dashed; hence, this weight pair
would not be found by any optimization algorithm based on a single
linear utility function. This point would be found, however, by a two-phase
algorithm minimizing driving time on the $a$-to-$d$ path and energy
consumption on the $d$-to-$f$ path.}
\label{fig:plot}
\end{figure}

We note, in addition, that in combining weights in this two-phase manner,
we are able to find Pareto-optimal scores that could not be found in any
optimization using a single linear utility function. That is, we can
find Pareto-optimal scores for paths from $s$ to $t$ that are not on the
convex hull of $(x,y)$ scores.
(See Figure~\ref{fig:plot}.)

Let us analyze the running time of this algorithm. Suppose, first,
that there are no negative-weight edges. In this case, we can use
Dijkstra's algorithm to compute each $T^{\rm out}_{s,i}$ and $T^{\rm in}_{t,j}$;
hence, these steps run in $O(c(n\log n+m))$ time.
If, on the other hand, there are negative-weight edges, but no
negative cycles, in $G$, then
we use a Bellman-Ford
algorithm to compute each $T^{\rm out}_{s,i}$ and $T^{\rm in}_{t,j}$;
hence, these steps run in $O(cnm)$ time in this case.
Then, computing all the $(i,j)^v_{i,j}$ pairs and choosing an optimal 
such pair takes $O(c^2n)$ time.
Thus, if $c$ is a fixed
constant independent of $n$ and $m$, then this algorithm runs
in $O(n\log n + m)$ time if there are no negative-weight edges and in 
$O(nm)$ time otherwise.
Note that these running times are asymptotically the same as that of
computing an optimal path for a single traversal mode.

In the context of finding electric vehicle routes,
each preference pair, $(\alpha_i,\beta_i)$,
corresponds to a driving style, such as ``minimize driving time,''
``minimize energy consumption,'' or ``minimize a weighted combination of driving
time and energy consumption.''
In addition, the path that achieves the chosen optimal pair,
$(x,y)^v_{i,j}$, is simple to implement for the driver of an electric
vehicle.
He or she simply needs to drive according to driving style~$i$ 
from $s$ to $v$, that is, using the path in $T^{\rm out}_{s,i}$, 
and then switch to drive according to driving style~$j$ from $v$ to $t$,
that is, using the path in $T^{\rm in}_{t,j}$.

\section{Including Charging Stations}
The above two-phase path finding algorithm can be used in the context
of negative-weight edges (e.g., where regenerative braking charges
the battery, provided we add the capacity constraints
as discussed in Section~\ref{pseudo_polynomial_sec}).
In this case, assuming there are no negative-weight
cycles, we could use the Bellman-Ford
algorithm to compute the optimal paths, requiring an $O(cnm+c^2n)$
running time.

If the charging stations themselves are the only places
in the network that provide negative energy consumption,
then we can achieve a potentially better algorithm for finding good paths.
In this case, we consider $s$ and $t$ to themselves 
to be charging stations, and
we let $d$ be the number of charging stations in the network.
Moreover, in this case, we assume the user is interested in the
shortest duration path
from $s$ to $t$ that can be achieved with a given battery capacity,
which starts out fully charged.
Also, we assume here that the user fully charges the battery at each 
charging station at which he or she stops.

With the algorithm we discuss in this section,
we can design a long route for 
an electric vehicle that starts at $s$,
and includes several charging stations,
fully charging the vehicle at each one along the way, and finally
going to $t$, such that we 
implement a different two-phase path between each pair of charging
stations along the way.
\begin{enumerate}
\item
For each charging station, $z$, and
each traversal mode, $(\alpha_i,\beta_i)$, use the Dijkstra-type algorithm of
Section~\ref{sec:modes} to find the tree, $T^{\rm out}_{z,i}$, that is the
union, for all $v$ in $G$,
of the optimal $z$-to-$v$ paths in $G$ for the traversal 
$(\alpha_i,\beta_i)$.
With each node, $v$, store the bicriterion weight, $(x,y)^{z,{\rm out}}_{i}$,
of the $z$-to-$v$ path in $T^{\rm out}_{z,i}$.

\item
For each charging station, $z$, and each
traversal mode, $(\alpha_j,\beta_j)$, use the (reverse) Dijkstra-type 
algorithm of
Section~\ref{sec:modes} to find the tree, $T^{\rm in}_{z,j}$, that is the
union, for all $v$ in $G$,
of the optimal $v$-to-$z$ paths in $G$ for the traversal 
$(\alpha_j,\beta_j)$.
With each node, $v$, store the bicriterion weight, $(x,y)^{z,{\rm in}}_{j}$,
of the $v$-to-$z$ path in $T^{\rm in}_{z,j}$.
\item
For each pair of charging stations, $u$ and $w$, and,
for each node $v$ in $G$, and each pair of indices
$i,j=1,2,\ldots,c$, 
compute the two-phase score,
\[
(x,y)^{u,v,w}_{i,j} = (x,y)^{u,{\rm out}}_i + (x,y)^{w,{\rm in}}_j,
\]
where ``$+$'' is component-wise addition.
\item
For each pair of charging stations, $u$ and $w$,
search all the $(x,y)^{u,v,w}_{i,j}$ values to find an optimal pair according
to the user's desired goals, to go from $u$ to $w$,
such as $x\le X$ and $y\le Y$ for given values of $X$ and $Y$.
Create a ``super edge,'' $e$, from $u$ to $w$, and label it with this
$(x,y)$ weight.
\item
Create a graph, $G'$, whose vertices are charging stations and whose
edges are the super edges created in the previous step.
For each such super edge, $e$, with weight, $(x,y)$, replace this
weight with the weight
\[
w(e)= x + {\rm charge}(C-y),
\]
where ${\rm charge}(E)$ is the time needed to charge the
battery to add $E$ units of energy capacity (and recall that $C$ is
the capacity of the battery).
\item
Use Dijkstra's algorithm to find a shortest duration path from $s$ to
$t$ in $G'$.
\end{enumerate}

\begin{figure}[htb]
\begin{center}
\includegraphics[width=3.3in, trim = 1.0in 0.75in 1.75in 0.25in, clip]{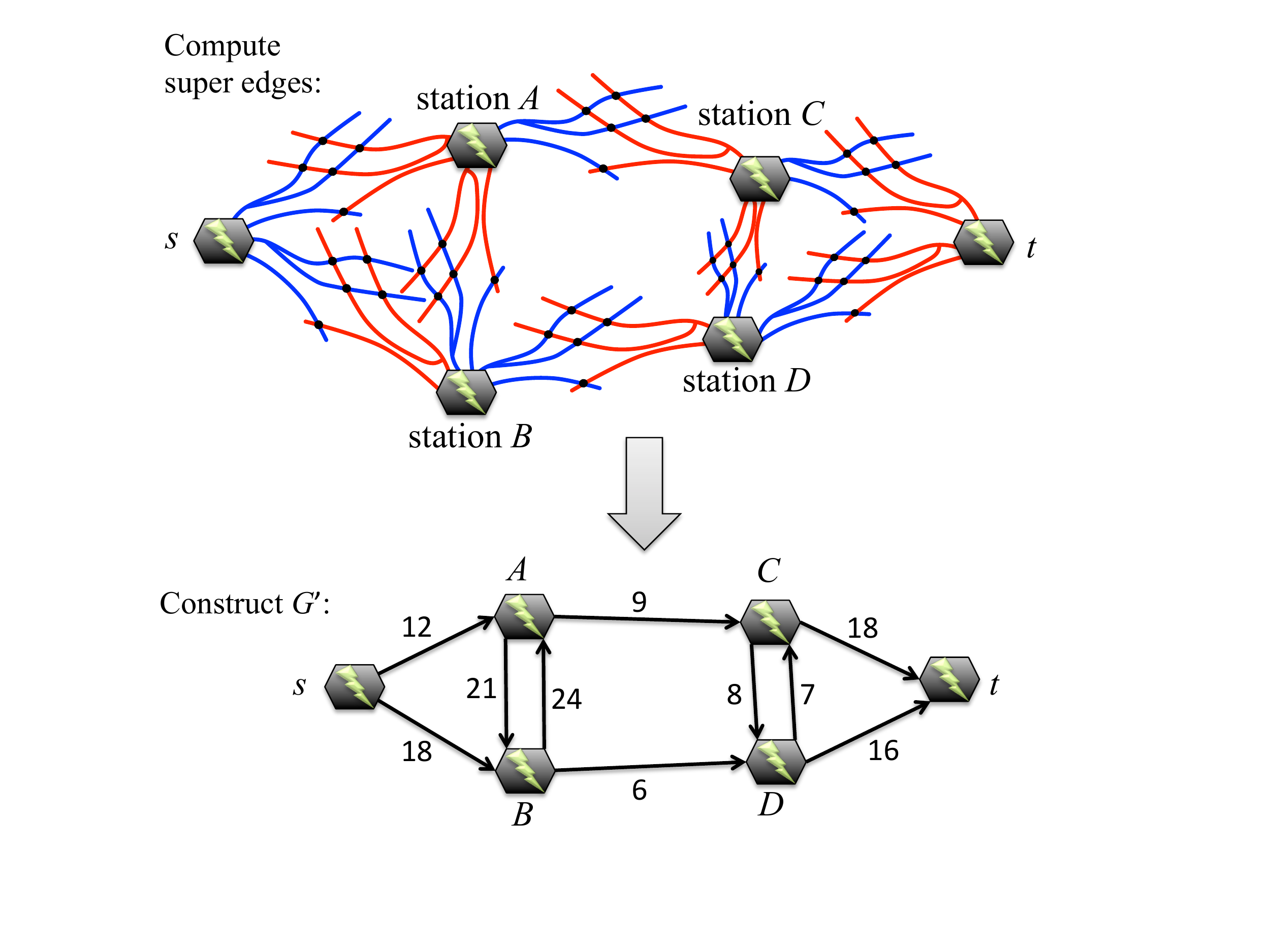}
\end{center}
\vspace*{-18pt}
\caption{An illustration of the algorithm for incorporating charging stations.
We consider $s$ and $t$ to be stations, then run the two-phase optimization
algorithm between all the stations. This gives us the graph, $G'$, where 
edge weights are now just driving time, since we know at this point which
stations can be driven between without depleting the battery (and we
always fully charge the battery at each charging station). Once we have the
graph, $G'$, we then do one more call to Dijkstra's algorithm to find
the shortest path from $s$ to $t$.
}
\label{fig:stations}
\end{figure}

We illustrate this algorithm in Figure~\ref{fig:stations}.

Incidentally, if there are negative-weight edges in the graph, but no
negative-weight cycles (ignoring charging stations), then we would
substitute the Dijkstra-type algorithms used in Steps~1 and~2 for 
Bellman-Ford-type algorithms.

Let us analyze 
the running time of this algorithm.
To compute all the trees of the form 
$T^{z,{\rm out}}_i$
and
$T^{z,{\rm in}}_j$, using Dijkstra's algorithm, takes $O(cd(n\log n+m))$.
The time to compute the optimal $(x,y)$ value for each super edge is
$O(c^2d^2n)$, but in practice we only need to consider each pair of
charging stations, $u$ and $w$, such that $w$ is reachable from $u$
with a fully charged battery. So the $d^2$ term in this bound might
be overly pessimistic.
Finally, the final Dijkstra's algorithm takes at most $O(d^2)$ 
time, but this is dominated by the running times of the other steps.
So the total running time of this algorithm is at most 
$O(c^2d^2n + cd(n\log n+m))$, assuming no negative-weight edges (other
than charging stations).
Note that
if $c$ and $d$ are fixed constants independent of $n$ and $m$, then
this running time is $O(n\log n + m)$, which is asymptotically the
same as doing a single Dijkstra-like computation with a single-phase
optimization criterion.

If there are negative-weight edges, but no negative-weight cycles, then 
replacing the Dijkstra-type algorithms in Steps~1 and~2 with Bellman-Ford-type
algorithms increases the running time 
to be $O(c^2d^2n + cdnm)$, which becomes asymptotically
equal to that of a single Bellman-Ford-type
computation, i.e., $O(nm)$, if $c$ and $d$ are fixed constants.

\section{Experiments}

\newcolumntype{R}{>{\centering\arraybackslash}X}
\renewcommand{\arraystretch}{1.2}

To empirically
measure the performance of our algorithms, we tested them
using road networks for several U.S. states from the TIGER/Line
data sets~\cite{tiger}, as prepared for the 9\textsuperscript{th} DIMACS
Implementation Challenge~\cite{dimacs}.
These road networks
are undirected, with each edge (road segment)
characterized as belonging to one of
four general classes: highway, primary major road, secondary major road, or
local road. 
For each
road segment of a given class, we consider $c=3$ different driving styles
for traversing an edge of that class, allowing for three different speeds
at which it can be traveled, in order to capture both lower and upper
speed limits inherent to all roads of a certain class. We derived these speeds
based on the guidelines presented in the road design manual for the state of
Florida~\cite{greenbook} (the ``Florida greenbook''). For these speed values, see
Table~\ref{driving_params_t}.

\begin{table}[hbt]
  \begin{tabularx}{0.48\textwidth}{|R|c|c|c|}
    \hline \multicolumn{2}{|c|}{\multirow{2}{*}{Road type}} & Speed & Energy Consumption \\
           \multicolumn{2}{|c|}{\multirow{2}{*}{}} & {[}mph{]} & {[}Wh / mile{]} \\
    \hline \multirow{3}{*}{Highway} & fast & 70 & 378 \\
    \cline{2-4} & moderate & 60 & 329 \\
    \cline{2-4} & slow & 50 & 291 \\
    \hline
    \multirow{3}{*}{\parbox[t]{1cm}{Primary main \\ road}} & fast & 70 & 378 \\
    \cline{2-4} & moderate & 55 & 308 \\
    \cline{2-4} & slow & 40 & 258 \\
    \hline \multirow{3}{*}{\parbox[t]{1cm}{Secondary main \\ road}} & fast & 60 & 329 \\
    \cline{2-4} & moderate & 45 & 275 \\
    \cline{2-4} & slow & 35 & 221 \\
    \hline \multirow{3}{*}{\parbox[t]{1cm}{Local \\ road}} & fast & 30 & 202 \\
    \cline{2-4} & moderate & 25 & 199 \\
    \cline{2-4} & slow & 20 & 197 \\
    \hline
  \end{tabularx}

  \caption{Driving parameters.}
  \label{driving_params_t}
\end{table}

Although our algorithms can accommodate elevation changes and even the
negative energy consumption that comes from regenerative braking,
the data sets in the TIGER/Line collection 
do not include elevation information; hence, for the sake of simplicity,
we assumed in our tests that all roads lie on a flat surface. 
Extending our testing regime to include elevation data would change some
of the weight pairs on some edges in hilly terrains, and would allow for 
including the second-order effect of elevation, but 
it would not significantly change the results for reasonably flat terrains.

Moreover, the main goal of our tests was to determine the effectiveness of
the two-phase strategy, for which the TIGER/Line data sets were sufficient.
In particular,
in order to estimate energy consumption for each edge segment, we
used the provided edge length and estimated energy consumption based on the
data for the Tesla Model S with 85 kWh battery~\cite{tesla,tesla2}
and air conditioning / heating turned on (see also Figure~\ref{fig-tesla-mph}).
The speed/energy consumption combinations are shown in Table~\ref{driving_params_t}.
For the two-phase algorithm from Section~\ref{two_phase_sec}, we considered
three driving styles:
\begin{itemize}
  \item emphasize smaller driving time
  \item emphasize smaller energy consumption
  \item balance energy consumption and driving time.
\end{itemize}
The preference pairs characterizing such paths are shown in Table~\ref{path_type_t}.

\begin{table}[h]
  \begin{tabularx}{0.48\textwidth}{|R|c|c|}
    \hline Path type & $\alpha$ (time coeff.) & $\beta$ (energy coeff.) \\
    \hline Fast & 0.8 & 0.2 \\
    \hline Balanced & 0.5 & 0.5 \\
    \hline Energy-saving & 0.2 & 0.8 \\
    \hline
  \end{tabularx}

  \caption{Path types.}
  \label{path_type_t}
\end{table}

\begin{table}[h!]
  Rhode Island ($n=53658$, $m=69213$):

  \begin{tabularx}{0.48\textwidth}{|c|c|c|R|c|}
    \hline Capacity & \multicolumn{2}{c|}{Reachable} & \multicolumn{2}{c|}{Two-phase algorithm} \\
    \cline{2-5} {[}Wh{]} & Nodes & \% $n$ & Reachability & Longer \% \\
    \hline 1000 & 2291 & 4.27 \% & 100 \% & 0.36 \% \\
    \hline 2000 & 3580 & 6.67 \% & 100 \% & 0.37 \% \\
    \hline 4000 & 9824 & 18.31 \% & 99.90 \% & 1.81 \% \\
    \hline 6000 & 23482 & 43.76 \% & 99.40 \% & 2.33 \% \\
    \hline 8000 & 44815 & 83.52 \% & 99.69 \% & 3.07 \% \\
    \hline
  \end{tabularx}

  \vspace{0.2in}
  Alaska ($n=69082$, $m=78100$):

  \begin{tabularx}{0.48\textwidth}{|c|c|c|R|c|}
    \hline Capacity & \multicolumn{2}{c|}{Reachable} & \multicolumn{2}{c|}{Two-phase algorithm} \\
    \cline{2-5} {[}Wh{]} & Nodes & \% $n$ & Reachability & Longer \% \\
    \hline 1000 & 2824 & 4.09 \% & 100 \% & 0.29 \% \\
    \hline 2000 & 7837 & 11.34 \% & 100 \% & 0.18 \% \\
    \hline 4000 & 9497 & 13.75 \% & 99.99 \% & 0.02 \% \\
    \hline 6000 & 11306 & 16.37 \% & 99.85 \% & 0.35 \% \\
    \hline 8000 & 12129 & 17.56 \% & 99.99 \% & 0.25 \% \\
    \hline 10000 & 13335 & 19.30 \% & 99.50 \% & 0.80 \% \\
    \hline 12000 & 17658 & 25.56 \% & 99.56 \% & 2.36 \% \\
    \hline
  \end{tabularx}

  \vspace{0.2in}
  Delaware ($n=49109$, $m=60512$):

  \begin{tabularx}{0.48\textwidth}{|c|c|c|R|c|}
    \hline Capacity & \multicolumn{2}{c|}{Reachable} & \multicolumn{2}{c|}{Two-phase algorithm} \\
    \cline{2-5} {[}Wh{]} & Nodes & \% $n$ & Reachability & Longer \% \\
    \hline 1000 & 3970 & 8.08 \% & 100 \% & 0.53 \% \\
    \hline 2000 & 12249 & 24.94 \% & 100 \% & 1.43 \% \\
    \hline 4000 & 18154 & 36.97 \% & 99.99 \% & 0.09 \% \\
    \hline 6000 & 19875 & 40.47 \% & 99.98 \% & 0.18 \% \\
    \hline 8000 & 21252 & 43.28 \% & 99.98 \% & 0.10 \% \\
    \hline 10000 & 23113 & 47.06 \% & 99.84 \% & 0.13 \% \\
    \hline 12000 & 26656 & 54.28 \% & 99.87 \% & 0.28 \% \\
    \hline 14000 & 28783 & 58.61 \% & 99.87 \% & 0.37 \% \\
    \hline 16000 & 31381 & 63.90 \% & 99.80 \% & 0.29 \% \\
    \hline
  \end{tabularx}

  \vspace{0.2in}
  District of Columbia ($n=9559$, $m=14909$):

  \begin{tabularx}{0.48\textwidth}{|c|c|c|R|c|}
    \hline Capacity & \multicolumn{2}{c|}{Reachable} & \multicolumn{2}{c|}{Two-phase algorithm} \\
    \cline{2-5} {[}Wh{]} & Nodes & \% $n$ & Reachability & Longer \% \\
    \hline 1000 & 3370 & 35.25 \% & 99.97 \% & 3.20 \% \\
    \hline 2000 & 8353 & 87.39 \% & 99.96 \% & 4.74 \% \\
    \hline 4000 & 9522 & 99.61 \% & 100 \% & 0.76 \% \\
    \hline
  \end{tabularx}

  \caption{Quality of the two-phase algorithm. Here, we use $n$ to denote
the number of vertices and $m$ to denote the number of edges in the underlying graph.}
  \label{quality_t}
\end{table}

\begin{figure}[h]
  \begin{center}
    \includegraphics[width=3.4in,trim=0.5in 0.5in 0.2in 0.8in,clip]{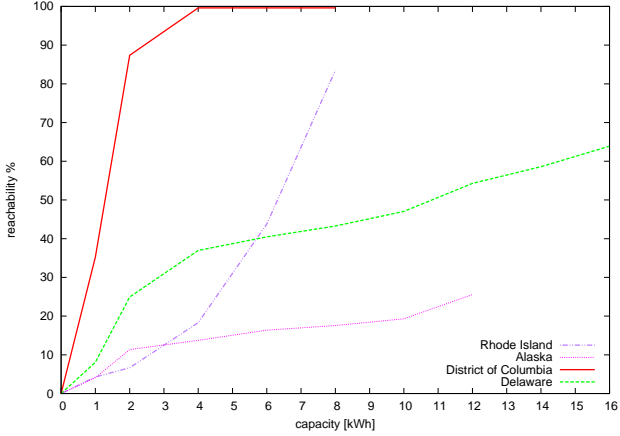}
  \end{center}

  \caption{Optimal reachability for small capacities.}
  \label{reachability_time_fig}
\end{figure}

\subsection{Quality of Paths}

As we argue above, the real-world goal of people driving electric
vehicles is to find a path that leads to the destination in the smallest amount
of time while ensuring that the battery stays at least partially charged at all
points along the way~\cite{Franke201356,APPS474,GrahamRowe2012140}. 
To measure the quality of the two-phase
bicriterion algorithm of Section~\ref{two_phase_sec}, we compared the paths it
returns against the optimal paths (that arrive at reachable destinations in
shortest time) found by the pseudo-polynomial time algorithm
of Section~\ref{pseudo_polynomial_sec}, where we set $N$ to be the capacity
(in Wh)
of the battery.

Due to the time complexity needed for finding optimal paths
using the vertex-labeling algorithm, we were only able to
compare the two algorithms on smaller graphs (with $n\leq 100000$), representing
small states (like Rhode Island, Delaware or the District of Columbia) or
large states with sparse road network (Alaska). 
In addition, 
due to time constraints imposed by the slow running time of the vertex-labeling
algorithm, we also did not consider placing charging stations in the graphs
for these comparison tests.

The results are shown in
Table~\ref{quality_t} and Figure~\ref{reachability_time_fig},
comparing the paths found by our algorithm
with the optimal paths found by the vertex-labeling algorithm. 
Due to high running times for the vertex-labeling algorithm (which depend
in a pseudo-polynomial fashion on battery capacity), we restricted
battery capacity to values much smaller than the actual 60 kWh (or 85 kWh) for
the Tesla Model S. These capacities are shown in the first column
of Table~\ref{quality_t}.
In the second and third column, we show the number of nodes reachable by
the vertex-labeling algorithm, 
both in absolute numbers and as a percentage of all
nodes in the network. The next column shows the percentage of the (optimally)
reachable nodes that can be reached by the two-phase algorithm. The final
column depicts the average slowdown of the paths computed by the
two-phase algorithm relative to the optimal paths.

\begin{table}
  California ($n=1613325$, $m=1989149$):

  \begin{tabularx}{0.48\textwidth}{|c|R|c|R|}
    \hline Capacity {[}Wh{]} & Chargers & Reachability & Time {[}s{]} \\
    \hline 60000 & 0 & 55.2 \% & 12.80 \\
    \hline 60000 & 1 & 56.2 \% & 24.63 \\
    \hline 60000 & 2 & 56.2 \% & 33.37 \\
    \hline 60000 & 3 & 95.3 \% & 53.98 \\
    \hline 60000 & 4 & 96.2 \% & 73.04 \\
    \hline 60000 & 5 & 97.6 \% & 91.39 \\
    \hline 85000 & 0 & 70.7 \% & 21.38 \\
    \hline 85000 & 1 & 77.0 \% & 28.26 \\
    \hline 85000 & 2 & 98.3 \% & 43.01 \\
    \hline
  \end{tabularx}

  \vspace{0.08in}
  Alaska ($n=69082$, $m=78100$):

  \begin{tabularx}{0.48\textwidth}{|c|R|c|R|}
    \hline Capacity {[}Wh{]} & Chargers & Reachability & Time {[}s{]} \\
    \hline 60000 & 0 & 29.2 \% & 0.48 \\
    \hline 60000 & 2 & 39.5 \% & 1.01 \\
    \hline 60000 & 5 & 40.8 \% & 2.03 \\
    \hline 60000 & 13 & 40.9 \% & 6.94 \\
    \hline 60000 & 14 & 43.3 \% & 7.70 \\
    \hline 60000 & 15 & 47.7 \% & 8.59 \\
    \hline 85000 & 0 & 43.6 \% & 0.48 \\
    \hline 85000 & 2 & 47.6 \% & 1.11 \\
    \hline 85000 & 13 & 47.7 \% & 7.86 \\
    \hline 85000 & 15 & 47.8 \% & 9.84 \\
    \hline
  \end{tabularx}

  \vspace{0.08in}
  Montana ($n=547028$, $m=670443$):

  \begin{tabularx}{0.48\textwidth}{|c|R|c|R|}
    \hline Capacity {[}Wh{]} & Chargers & Reachability & Time {[}s{]} \\
    \hline 60000 & 0 & 88.3 \% & 9.90 \\
    \hline 60000 & 1 & 88.4 \% & 12.83 \\
    \hline 60000 & 2 & 96.1 \% & 18.46 \\
    \hline 60000 & 3 & 96.7 \% & 22.39 \\
    \hline 60000 & 6 & 97.5 \% & 39.00 \\
    \hline 60000 & 7 & 97.9 \% & 57.42 \\
    \hline 85000 & 0 & 97.0 \% & 9.93 \\
    \hline 85000 & 1 & 97.3 \% & 14.01 \\
    \hline 85000 & 2 & 98.2 \% & 20.33 \\
    \hline
  \end{tabularx}

  \vspace{0.08in}
  Texas ($n=2073870$, $m=2584159$):

  \begin{tabularx}{0.48\textwidth}{|c|R|c|R|}
    \hline Capacity {[}Wh{]} & Chargers & Reachability & Time {[}s{]} \\
    \hline 60000 & 0 & 47.2 \% & 22.83 \\
    \hline 60000 & 1 & 49.8 \% & 26.55 \\
    \hline 60000 & 2 & 56.1 \% & 49.86 \\
    \hline 60000 & 3 & 57.9 \% & 64.33 \\
    \hline 60000 & 4 & 58.4 \% & 89.50 \\
    \hline 60000 & 5 & 69.2 \% & 113.34 \\
    \hline 60000 & 7 & 69.3 \% & 154.51 \\
    \hline 60000 & 9 & 71.2 \% & 190.46 \\
    \hline 85000 & 0 & 68.7 \% & 28.73 \\
    \hline 85000 & 1 & 75.1 \% & 35.10 \\
    \hline 85000 & 2 & 80.6 \% & 58.75 \\
    \hline 85000 & 3 & 82.4 \% & 86.36 \\
    \hline 85000 & 5 & 94.9 \% & 138.59 \\
    \hline
  \end{tabularx}

  \caption{Performance of the two-phase algorithm.}
  \label{performance_t}
\end{table}

\begin{table}
  Nevada ($n=261155$, $m=311043$):

  \begin{tabularx}{0.48\textwidth}{|c|R|c|R|}
    \hline Capacity {[}Wh{]} & Chargers & Reachability & Time {[}s{]} \\
    \hline 60000 & 0 & 55.7 \% & 2.87 \\
    \hline 60000 & 1 & 63.2 \% & 4.81 \\
    \hline 60000 & 2 & 67.9 \% & 6.43 \\
    \hline 60000 & 4 & 81.9 \% & 11.62 \\
    \hline 60000 & 10 & 92.6 \% & 34.06 \\
    \hline 85000 & 0 & 80.6 \% & 3.44 \\
    \hline 85000 & 1 & 92.6 \% & 5.99 \\
    \hline
  \end{tabularx}

  \caption{Performance of the two-phase algorithm (continued).}
  \label{performance_t2}
\end{table}

\subsection{Performance}

\begin{figure}[h]
  \begin{center}
    \includegraphics[width=3.4in,trim=0.5in 0.5in 0.2in 0.8in,clip]{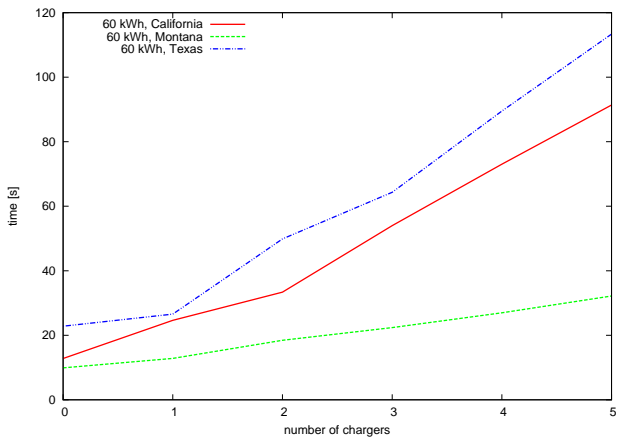}
  \end{center}
  \caption{Dependence of running time on the number of charging stations.}
  \label{time_fig}
\end{figure}

\begin{figure}
  \begin{center}
    \includegraphics[width=3.4in,trim=0.5in 0.5in 0.2in 0.8in,clip]{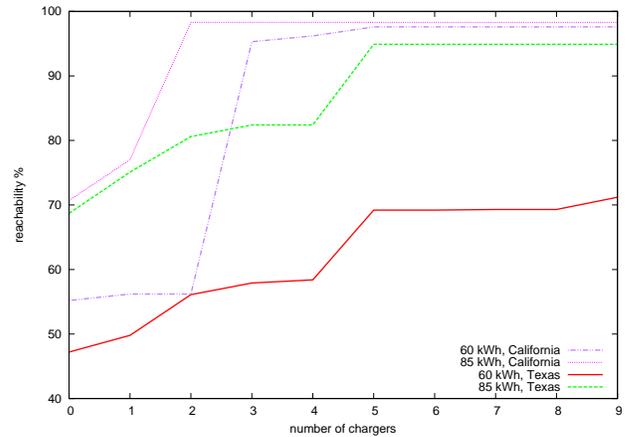}
  \end{center}
  \caption{Dependence of reachability on the number of charging stations.}
  \label{reachability_charge_fig}
\end{figure}

As mentioned above, due to the extremely large running time of the optimal
pseudo-polynomial algorithm (for the largest instances our runs
exceeded 24 hours), we
were forced to restrict our qualitative 
testing to road networks of small states, and use
unrealistically small battery capacities. In this subsection, we focus on the
performance of the two-phase algorithm, which, thanks to its superior time
complexity, allows us to meet the following goals:
\begin{itemize}
  \item Use actual capacities of Tesla Model S (60/85 kWh).
  \item Include charging stations.
  \item Test the algorithm on larger graphs.
\end{itemize}
Under the above assumptions, we measured the running time of the algorithm,
as well as estimated the reachability percentage (measured as a ratio of
feasible paths between pairs of randomly chosen vertices and the total
number of pairs tested; in each case, we tested 1000 pairs). The results are
summarized in Table~\ref{performance_t}, Table~\ref{performance_t2},
Figure~\ref{time_fig} and Figure~\ref{reachability_charge_fig}.
The times shown in the last column
is the average duration of a single execution of the two-phase algorithm.
Charging stations were placed at randomly selected vertices. Only instances
that actually increased reachability are shown.

\newpage

\subsection{Discussion}

Tests were implemented in \textsc{C++} and carried out on a PC with a
2.2 GHz CPU, 1066 MHz
bus, and 4 GB RAM running Linux.
It is evident that the two-phase algorithm finds paths to almost all reachable
destinations, with the paths being only slightly slower
(taking more time) that the optimal ones.

Our results were obtained using the following procedure:
for each state, we randomly chose a starting position and 1000 destinations.
It gave us 1000 origin-destination pairs, on which we then tested the
algorithms described above. The resolution of our algorithms was: seconds
(for time) and Wh (for energy).

Our implementation of the two-phase algorithm is straightforward. We did not
optimize it for running time and we deliberately ran it
on a relatively old PC, and, admittedly, this shows in the results.
Even then, the algorithm was able to compute paths within several dozens of
seconds. Since the number of charging stations is the main factor in running
time, one optimization would be to precompute best paths between all pairs of
charging stations (which is feasible, as the number of charging stations is
small and they are fixed features of a road network). The running time of the
algorithm would then be reduced to the case of no charging stations.
As the main component of our procedure is the Dijkstra's shortest path algorithm,
another straightforward improvement would be to incorporate
some of existing approaches~\cite{DBLP:conf/dfg/DellingSSW09,DBLP:conf/wea/DellingGPW11}
aimed at speeding up Dijkstra's algorithm.

\section{Conclusion}
We have presented a two-phase approach for finding good paths in
bicriterion networks, and we have demonstrated that our algorithms are both
fast and effective for finding good routes for electric vehicles.
In particular, we have shown empirically that the paths found by the two-phase
algorithm can identify over 99\% of the vertices reachable in a road network
by some energy-efficient algorithm, while being only
slightly longer on average than paths found by the inefficient vertex-labeling
algorithm.
Moreover, we believe that two-phase
are easier for people to follow, since, in addition
to the route they plan to take, they only need to remember two
different driving styles and the point in the route where they transition from
the first driving style to the second. Of course, if $k$ charging stations
are involved, it may require $2k-1$ style transitions. This is usually not a
problem, since common trips tend to use a small number of charging station
located far away.

As possible future work, it would be interesting to test the two-phase
approach for finding good delivery routes for electric vehicles that have
multiple destinations.

\subsection*{Acknowledgments}
This work was supported in part by the NSF, under grant 1011840
and 1228639,
and by the Office
of Naval Research, under grant N00014-08-1-1015.
We would like to thank David Eppstein and Amelia Regan for several helpful
communications regarding the topics of this paper.

\begin{flushleft}
\bibliographystyle{abbrv}
\bibliography{refs}
\end{flushleft}

\end{document}